\documentclass[seceq,preprint]{ptptex}
\usepackage{wrapft}
\usepackage{graphicx}

\newcommand{\bra}[1]{\langle {#1} |}     
\newcommand{\ket}[1]{| {#1} \rangle}     
\newcommand{\rket}[1]{| {#1} )}     
\newcommand{\rdket}[1]{|\!| {#1} )}     
\newcommand{\dket}[1]{|\!| {#1} \rangle}     
\newcommand{\wtilde}[1]{\widetilde{#1}} 

\def\<{\langle}
\def\>{\rangle}
\def\bsub{\begin{subequations}}
\def\esub{\end{subequations}}
\def\beqn{\begin{eqnarray}}
\def\eeqn{\end{eqnarray}}
\def\b{\begin{equation}}

\title{
Boson Realization of the $su(3)$-Algebra. I
}
\subtitle{
Schwinger Representation for the Lipkin Model
}    

\author{
Constan\c{c}a {\sc Provid\^encia},$^{1}$
Jo\~ao da {\sc Provid\^encia},$^{1}$\\
Yasuhiko {\sc Tsue}$^{2}$ 
and Masatoshi {\sc Yamamura}$^{3}$
}

\inst{
$^{1}$Departamento de Fisica, Universidade de Coimbra, 3004-516 Coimbra, 
Portugal\\
$^2$Physics Division, Faculty of Science, Kochi University, Kochi 780-8520, 
Japan\\
$^{3}$Faculty of Engineering, Kansai University, Suita 564-8680, Japan
}

\recdate{
\today
}

\abst{
Following a general form for the Schwinger boson representation of the 
$su(M+1)$ Lipkin model presented in the previous paper, three types of 
the orthogonal sets characterizing the $su(3)$-algebra are proposed. 
In these three, third is presented in Appendix. 
The intrinsic state is specified by two quantum numbers and two excited 
state generating operators play a central role. 
}
\begin{document}
\maketitle

\section{Introduction}

It is well known that the Lipkin model\cite{1} has played a crucial role 
in microscopic studies of collective motion observed in nuclei. 
The original Lipkin model is based on the $su(2)$-algebra and, with the aid of 
this model, we can understand schematically collective motion 
based on superposition of particle-hole pair excitations. 
However, in order to serve the studies of coupling of collective and 
non-collective degree of freedom, which is one of the most fundamental 
problems in nuclear theory, the $su(2)$-Lipkin model must be generalized 
and the simplest generalization may be the $su(3)$-algebraic model.\cite{2} 
On the other hand, this model has been also served as the schematic 
understanding of finite temperature effects.\cite{3} 
In this case, the use of the free vacuum is helpless. 
The above situations are mentioned in detail in Ref.\citen{4}, 
which is, hereafter, referred to as (A).

A main aim of this paper, (I), is to formulate the Schwinger boson 
representation for the Lipkin model, which is presented in (A), for the case 
of the $su(3)$-algebra in complete form. 
In (A), only the surface is sketched. 
The $su(3)$-algebra consists of eight generators, in which some generators 
determine the intrinsic state and some of others play a role of the excited 
state generating operators. 
With the use of these operators, we can determine the orthogonal set 
of the $su(3)$-algebra completely. 
Depending on the properties of the excited state generating operators, 
three forms are presented. 
In \S 4, \S 5 and Appendix, the details are discussed. 
As was already treated in (A), the present $su(3)$-algebra is described 
in terms of six kinds of boson operators : 
two sets of vector types. 
But, we know that the orthogonal set of the $su(3)$-algebra is specified 
by five quantum numbers, and the remaining one is used for the 
$su(1,1)$-algebra.

In the next section, the $su(3)$-algebra in the Schwinger representation 
is recapitulated in terms of the form presented in (A). Section 3 is devoted 
to constructing the intrinsic state specified by two quantum numbers. 
In \S 4, the first type of the orthogonal set is given by regarding 
the two excited state generating operators as tensor operators with 
rank 1/2 (spinor). 
The main part of \S 5 is to construct the second type in terms of 
successive operations of the operators related to the 
excited state generating operatos on the intrinsic state. 
In \S 5, some characteristic points obtained in \S\S 3 and 4 are discussed. 
In Appendix, the third form is given in the form of successive operations 
of the excited state generating operators together with certain projection 
operators.

\section{The $su(3)$-algebra in the form suitable for the Lipkin model 
and its Schwinger representation}

As was recapitulated in (A), the $su(3)$-algebra is composed of 
eight generators, which were, in (A), denoted as 
$({\hat S}_2^1, {\hat S}_1^2, {\hat S}_1^1, {\hat S}_2^2, {\hat S}^1, 
{\hat S}^2, {\hat S}_1, {\hat S}_2)$. 
In this paper, we formulate the $su(3)$-algebra in a form, which 
may be suitable for the boson realization of the Lipkin model. 
Therefore, the notations different from the above are adopted:
\bsub\label{2-1}
\beqn
& &{\hat I}_+={\hat S}_2^1\ , \qquad {\hat I}_-={\hat S}_1^2 \ , 
\qquad {\hat I}_0=(1/2)({\hat S}_2^2-{\hat S}_1^1) \ , 
\label{2-1a}\\
& &{\hat M}_0=(1/2)({\hat S}_2^2+{\hat S}_1^1) \ , 
\label{2-1b}\\
& &{\hat D}_-^*={\hat S}^1 \ , \qquad
{\hat D}_+^*={\hat S}^2 \ , \qquad 
{\hat D}_-={\hat S}_1 \ , \qquad
{\hat D}_+={\hat S}_2 \ . 
\label{2-1c}
\eeqn
\esub
The above operators obey the following commutation relations: 
\bsub\label{2-2}
\beqn
& &[\ {\hat I}_+ \ , \ {\hat I}_-\ ]=2{\hat I}_0 \ , 
\qquad
[\ {\hat I}_0 \ , \ {\hat I}_\pm\ ]=\pm{\hat I}_\pm \ , 
\label{2-2a}\\
& &[\ {\hat I}_{\pm,0} \ , \ {\hat M}_0\ ]=0 \ , 
\label{2-2b}\\
& &[\ {\hat I}_{\pm} \ , \ {\hat D}_\pm^*\ ]=0 \ , \qquad
[\ {\hat I}_{\pm} \ , \ {\hat D}_\mp^*\ ]={\hat D}_{\pm}^* \ , \nonumber\\
& &[\ {\hat I}_{0} \ , \ {\hat D}_\pm^*\ ]=\pm (1/2){\hat D}_\pm^*\ , 
\label{2-2c}\\
& &[\ {\hat M}_{0} \ , \ {\hat D}_\pm^*\ ]=(3/2){\hat D}_\pm^*\ , 
\label{2-2d}\\
& &[\ {\hat D}_{+}^* \ , \ {\hat D}_-^*\ ]=0\ , \qquad
[\ {\hat D}_{\pm}^* \ , \ {\hat D}_\pm\ ]={\hat M}_0\pm {\hat I}_0\ , \qquad
[\ {\hat D}_{\pm}^* \ , \ {\hat D}_\mp\ ]={\hat I}_\pm\ .
\label{2-2e}
\eeqn
\esub
We can see in the relation (\ref{2-2a}) that the set $({\hat I}_{\pm,0})$ 
obeys the $su(2)$-algebra, and further, the set $({\hat D}_{\pm}^*)$ forms 
a spinor, i.e., a spherical tensor with rank 1/2 with respect to 
$({\hat I}_{\pm,0})$, in which $+$ and $-$ correspond to the $z$-components 
$+1/2$ and $-1/2$, respectively. 
We can see this fact in the relation (\ref{2-2c}). The Casimir operator 
${\hat \Gamma}_{su(3)}$ can be expressed in the form 
\beqn\label{2-3}
{\hat \Gamma}_{su(3)}&=&
{\hat S}^1{\hat S}_1 +{\hat S}_1{\hat S}^1+{\hat S}^2{\hat S}_2
+{\hat S}_2{\hat S}^2+{\hat S}_2^1{\hat S}_1^2+{\hat S}_1^2{\hat S}_2^1
\nonumber\\
& &+(2/3)\left[({\hat S}_1^1)^2-{\hat S}_1^1{\hat S}_2^2+
({\hat S}_2^2)^2\right]\ . 
\eeqn
The form (\ref{2-3}) can be reexpressed as 
\begin{equation}\label{2-4}
{\hat \Gamma}_{su(3)}=2\left[
{\hat {\mib D}}^*{\hat {\mib D}}+{\hat {\mib D}}{\hat {\mib D}}^*
+(1/3){\hat M}_0^2+{\hat {\mib I}}^2\right] \ , 
\end{equation}
\vspace{-0.5cm}
\bsub\label{2-5}
\beqn
& &{\hat {\mib D}}^*{\hat {\mib D}}=(1/2)({\hat D}_+^*{\hat D}_+
+{\hat D}_-^*{\hat D}_-) \ , \nonumber\\
& &{\hat {\mib D}}{\hat {\mib D}}^*=(1/2)({\hat D}_+{\hat D}_+^*
+{\hat D}_-{\hat D}_-^*) \ , 
\label{2-5a}\\
& &{\hat {\mib I}}^2={\hat I}_0^2+(1/2)({\hat I}_+{\hat I}_-
+{\hat I}_-{\hat I}_+) \ . 
\label{2-5b}
\eeqn
\esub

Associating the above $su(3)$-algebra, the $su(1,1)$-algebra 
also plays a central role in the present form. 
It is composed of three generators which were, in (A), denoted as 
$({\hat T}^1, {\hat T}_1, {\hat T}_1^1)$. 
In this paper, we will use the notation $({\wtilde T}_{\pm,0})$ which 
is defined as 
\b\label{2-6}
{\wtilde T}_+={\hat T}^1 \ , \qquad
{\wtilde T}_-={\hat T}_1 \ , \qquad 
{\wtilde T}_0=(1/2){\hat T}_1^1 \ . 
\end{equation}
The commutation relation and the Casimir operator are given as follows: 
\beqn
& &[\ {\wtilde T}_+ \ , \ {\wtilde T}_- \ ]=-2{\wtilde T}_0 \ , \qquad
[\ {\wtilde T}_0 \ , \ {\wtilde T}_\pm \ ]=\pm{\wtilde T}_\pm \ , 
\label{2-7}\\
& &{\hat \Gamma}_{su(1,1)}=2{\wtilde {\mib T}}^2 \ , 
\label{2-8}\\
& &{\wtilde {\mib T}}^2={\wtilde T}_0^2-(1/2)(
{\wtilde T}_+{\wtilde T}_-+{\wtilde T}_-{\wtilde T}_+) \ . 
\label{2-9}
\eeqn

A possible boson realization of the above two algebras is 
obtained in the framework of six kinds of boson operators 
$({\hat a}, {\hat a}^*)$, $({\hat b}, {\hat b}^*)$, 
$({\hat a}_{\pm}, {\hat a}_{\pm}^*)$ and 
$({\hat b}_{\pm}, {\hat b}_{\pm}^*)$. 
The detail can be found in (A). 
By changing the notations adopted in (A) to those suitable for the present 
form such as ${\hat a}^1\rightarrow {\hat a}$, 
${\hat a}_1\rightarrow {\hat a}_-$, ${\hat a}_2\rightarrow {\hat a}_+$, 
${\hat b}\rightarrow {\hat b}$, ${\hat b}_2^1\rightarrow {\hat b}_-$, 
${\hat b}_1^1\rightarrow {\hat b}_+$, 
we have the following expressions: 
\bsub\label{2-10}
\beqn
& &{\hat I}_+={\hat a}_+^*{\hat a}_- - {\hat b}_+^*{\hat b}_- \ , 
\qquad
{\hat I}_-={\hat a}_-^*{\hat a}_+ - {\hat b}_-^*{\hat b}_+\ , 
\nonumber\\
& &{\hat I}_0=(1/2)\left[({\hat a}_+^*{\hat a}_+ - {\hat a}_-^*{\hat a}_-) 
+({\hat b}_+^*{\hat b}_+ - {\hat b}_-^*{\hat b}_-)\right] \ , 
\label{2-10a}\\
& &{\hat M}_0=({\hat a}^*{\hat a}-{\hat b}^*{\hat b})+
(1/2)\left[({\hat a}_+^*{\hat a}_+ + {\hat a}_-^*{\hat a}_-) 
-({\hat b}_+^*{\hat b}_+ + {\hat b}_-^*{\hat b}_-)\right] \ , 
\label{2-10b}\\
& &{\hat D}_-^*={\hat a}_-^*{\hat b} + {\hat a}^*{\hat b}_+ \ , 
\qquad
{\hat D}_+^*={\hat a}_+^*{\hat b} + {\hat a}^*{\hat b}_-\ , 
\nonumber\\
& &{\hat D}_-={\hat b}^*{\hat a}_- + {\hat b}_+^*{\hat a} \ , 
\qquad
{\hat D}_+={\hat b}^*{\hat a}_+ + {\hat b}_-^*{\hat a}\ , 
\qquad\qquad\qquad\qquad\qquad
\label{2-10c}
\eeqn
\esub
\vspace{-0.8cm}
\beqn\label{2-11}
& &{\wtilde T}_+={\hat a}^*{\hat b}^* - {\hat a}_+^*{\hat b}_-^*
-{\hat a}_-^*{\hat b}_+^* \ , \qquad
{\wtilde T}_-={\hat b}{\hat a} - {\hat b}_-{\hat a}_+
-{\hat b}_+{\hat a}_- \ , \nonumber\\
& &{\wtilde T}_0=(1/2)\left[
({\hat a}^*{\hat a}+{\hat b}^*{\hat b})+({\hat a}_+^*{\hat a}_+
+{\hat b}_+^*{\hat b}_+)+({\hat a}_-^*{\hat a}_-+{\hat b}_-^*{\hat b}_-)
+3\right] \ . 
\eeqn

Calculation of the commutation relations for $({\hat I}_{\pm,0})$ 
defined in the relation (\ref{2-10a}) tells us that the bosons 
${\hat a}^*$ and ${\hat b}^*$ are scalars (rank $=0$) and 
$({\hat a}_-^*, {\hat a}_+^*)$ and $({\hat b}_-^*, {\hat b}_+^*)$ 
are spinors (rank $=1/2$). 
Further, it may be important to see that the representations 
(\ref{2-10}) and (\ref{2-11}) give us the relation 
\begin{equation}\label{2-12}
[\ \hbox{\rm any\ of\ the\ $su(3)$-generators}\ , \ 
\hbox{\rm any\ of\ the\ $su(1,1)$-generators}\ ]=0 \ . 
\end{equation}
The Casimir operator (\ref{2-4}) can be reexpressed in the form 
\beqn
& &{\hat \Gamma}_{su(3)}=2\left[
{\wtilde {\mib T}}^2-3/4+(1/3)({\wtilde R}_0)^2\right] \ , 
\label{2-13}\\
& &{\wtilde R}_0=(1/2)
\left[({\hat a}^*{\hat a}-{\hat b}^*{\hat b})-
({\hat a}_+^*{\hat a}_+-{\hat b}_+^*{\hat b}_+)
-({\hat a}_-^*{\hat a}_- -{\hat b}_-^*{\hat b}_-)\right] \ . 
\label{2-14}
\eeqn
In (A), ${\wtilde R}_0$ is denoted as ${\hat R}_0$. 
Naturally, we have 
\begin{equation}\label{2-15}
[\ {\wtilde R}_0 \ , \ \hbox{\rm any\ of\ the\ $su(3)$-\ and\ 
the\ $su(1,1)$-generators}\ ]=0 \ . 
\end{equation}
The proof of the relation (\ref{2-13}) is straightforward, but tedious. 
The above is an outline of the $su(3)$-algebra and its associating 
$su(1,1)$-algebra in the Schwinger representation suitable for the 
Lipkin model. 
The system under investigation is composed of six kinds of boson operators.

\section{Construction of the intrinsic state}

The present system consists of six kinds of boson operators and we 
can easily find that there exist totally six commuted operators: 
${\wtilde {\mib T}}^2$, ${\wtilde T}_0$, ${\wtilde R}_0$, 
${\hat {\mib I}}^2$, ${\hat I}_0$ and ${\hat M}_0$. 
Therefore, the eigenstates for the above operators provide a complete 
orthogonal set for the $su(3)$-algebra and its associating $su(1,1)$-algebra. 
For the task obtaining this orthogonal set, we, first, construct the 
intrinsic state introduced in (A).

As was discussed in (A), the intrinsic state, which we denote $\ket{m}$, 
should obey the following condition: 
\b\label{3-1}
{\hat D}_-\ket{m}={\hat D}_+\ket{m}={\hat I}_-\ket{m}=0\ , \qquad
{\wtilde T}_-\ket{m}=0 \ .
\end{equation}
In the notation adopted in (A), the condition (\ref{3-1}) can be expressed 
as ${\hat S}_1\ket{m}={\hat S}_2\ket{m}={\hat S}_1^2\ket{m}=0$ and 
${\hat T}_1\ket{m}=0$. 
Further, $\ket{m}$ should be an eigenstate for ${\wtilde T}_0$, 
${\wtilde R}_0$, ${\hat I}_0$ and ${\hat M}_0$. 
The state $\ket{m}$ which satisfies the above condition is 
easily obtained: 
\beqn\label{3-2}
\ket{m}&=&\ket{T,R}\nonumber\\
&=&\left(\sqrt{(T-3/2+R)!(T-3/2-R)!}\right)^{-1}
({\hat b}_-^*)^{T-3/2+R}({\hat b}^*)^{T-3/2-R}\ket{0} \ . 
\eeqn
It satisfies the eigenvalue equations 
\bsub\label{3-3}
\beqn
& &{\wtilde T}_0\ket{T,R}=T\ket{T,R} \ , 
\label{3-3a}\\
& &{\wtilde R}_0\ket{T,R}=R\ket{T,R} \ , 
\label{3-3b}\\
& &{\hat I}_0\ket{T,R}=-I^0\ket{T,R} \ , \qquad
I^0=(1/2)(T-3/2+R) \ , 
\label{3-3c}\\
& &{\hat M}_0\ket{T,R}=-M^0\ket{T,R} \ , \qquad
M^0=(1/2)(3(T-3/2)-R) \ . 
\label{3-3d}
\eeqn
\esub
Since the boson numbers are positive-integers, the form of ${\wtilde T}_0$ 
shown in the relation (\ref{2-11}) gives us 
\begin{equation}\label{3-4}
T=3/2\ , \ 2\ , \ 5/2\ , \ 3\ , \ \cdots \ . \quad (T\geq 3/2)
\end{equation}
Further, the exponents $(T-3/2+R)$ and $(T-3/2-R)$ appearing 
in the state (\ref{3-2}) should be positive-integers 
and we have 
\beqn\label{3-5}
& &R=-(T-3/2)\ , \ -(T-3/2)+1\ , \ \cdots , \ (T-3/2)-1\ , \ (T-3/2)\ . 
\nonumber\\
& &\quad\qquad\qquad\qquad (-(T-3/2) \leq R \leq T-3/2)
\eeqn
The relations (\ref{3-3c})$\sim$ (\ref{3-5}) give us 
\beqn
& &I^0=0\ , \ 1/2\ , \ 1\ , \ 3/2\ , \ \cdots \ , \ (T-3/2) \ , 
\quad (0 \leq I^0 \leq T-3/2) 
\label{3-6}\\
& &M^0=I^0\ , \ I^0+1\ , \ I^0+2\ , \ \cdots \ .
\quad (M^0 \geq I^0 ) 
\label{3-7}
\eeqn
The eigenvalues of ${\hat \Gamma}_{su(3)}$ and ${\hat \Gamma}_{su(1,1)}$ 
for the state $\ket{T,R}$ are expressed as 
\begin{equation}\label{3-8}
\hbox{\rm the\ eigenvalue\ of\ }
\begin{cases} {\hat \Gamma}_{su(3)}=2\left[
(T-3/2)(T+1/2)+(1/3)R^2\right] \ , \\
{\hat \Gamma}_{su(1,1)}=2T(T-1) \ . 
\end{cases}
\end{equation}
Instead of $(T,R)$, we can specify the state $\ket{m}$ in terms of 
$(M^0, I^0)$. 
In this specification, we have 
\begin{equation}\label{3-10}
\hbox{\rm the\ eigenvalue\ of\ }
\begin{cases} {\hat \Gamma}_{su(3)}=2\left[(1/3)M^0(M^0+3)
+I^0(I^0+1)\right] \ , \\
{\hat \Gamma}_{su(1,1)}=2\cdot(1/2)(M^0+I^0+3)\cdot(1/2)(M^0+I^0+1) \ . 
\end{cases}
\end{equation}
The quantum numbers $M^0$ and $I^0$ correspond to $\lambda$ and $\mu$ in 
the Elliott model in the relations $M^0=\lambda$ and $I^0=\mu$. 
The state $\ket{m}$ is also specified by $(T, I^0)$. 
In this case, we have 
\begin{equation}\label{3-12}
M^0=2(T-3/2)-I^0 \ , \qquad
R=2I^0-(T-3/2) \ . 
\end{equation}
This specification will play a central role in \S 4: 
$\ket{I^0,T}$.

\section{Structure of the excited states}

In (A), we mentioned that the excited states constructed on the intrinsic 
state $\ket{m}$ are composed in terms of the excited state generating 
operators ${\hat S}^1$, ${\hat S}^2$ and ${\hat S}_2^1$, in the 
present notations ${\hat D}_-^*$, ${\hat D}_+^*$ and ${\hat I}_+$. 
If including the $su(1,1)$-algebra, ${\wtilde T}_+$ is added. 
With the use of these operators, we construct the orthogonal set. 
First, we note two points. 
Since ${\hat I}_-\ket{I^0,T}=0$ and ${\hat I}_0\ket{I^0,T}=
-I^0\ket{I^0,T}$, a possible eigenstate of ${\hat {\mib I}}^2$ and 
${\hat I}_0$, which we denote as $\ket{I^0I_0^0;T}$, is given 
in the form 
\beqn\label{4-1}
\ket{I^0I_0^0;T}&=&
\sqrt{\frac{1}{(2I^0)!}\frac{(I^0-I_0^0)!}{(I^0+I_0^0)!}}
({\hat I}_+)^{I^0+I_0^0}\ket{I^0,T} \ , \nonumber\\
I^0_0&=&
-I^0 \ , \ -I^0+1\ , \ \cdots \ , I^0-1\ , \ I^0\ . \quad
(-I^0\leq I_0^0 \leq I^0)
\eeqn
The above is the first point. 
Next, we consider the second point. We have already mentioned that 
$({\hat D}_{\pm}^*)$ is a spherical tensor with rank 1/2, the $z$-components 
of which are specified by $\pm 1/2$. 
Then, a spherical tensor with rank $I^1$ and $z$-component $I_0^1$, 
${\hat D}_{I^1I_0^1}^*$, is obtained in the form 
\begin{equation}\label{4-2}
{\hat D}_{I^1I_0^1}^*=\left(\sqrt{(I^1+I_0^1)!(I^1-I_0^1)!}\right)^{-1}
({\hat D}_+^*)^{I^1+I_0^1}({\hat D}_-^*)^{I^1-I_0^1} \ . 
\end{equation}
We can prove the relation 
\beqn\label{4-3}
& &[\ {\hat I}_\pm \ , \ {\hat D}_{I^1I_0^1}^* \ ]=
\sqrt{(I^1\mp I_0^1)(I^1\pm I_0^1 +1)}{\hat D}_{I^1I_0^1\pm 1}^* \ , 
\nonumber\\
& &[\ {\hat I}_0 \ , \ {\hat D}_{I^1I_0^1}^* \ ]=
I_0^1{\hat D}_{I^1I_0^1}^* \ . 
\eeqn
Then, we define the following state: 
\begin{equation}\label{4-4}
\ket{I^1I^0,II_0;T}
=\sum_{I_0^1I_0^0}\bra{I^1I_0^1I^0I_0^0}II_0\rangle
{\hat D}_{I^1I_0^1}^*\ket{I^0I_0^0;T} \ . 
\end{equation}
Here, $\bra{I^1I_0^1I^0I_0^0}II_0\rangle$ denotes the Clebsch-Gordan 
coefficient. Clearly, the state $\ket{I^1I^0,II_0;T}$ is an eigenstate of 
${\wtilde {\mib T}}^2$, ${\wtilde T}_0$, ${\wtilde R}_0$, 
${\hat {\mib I}}^2$, ${\hat I}_0$ and ${\hat M}_0$: 
\begin{equation}\label{4-5}
\hbox{\rm the\ eigenvalue\ of\ }
\begin{cases} 
{\wtilde {\mib T}}^2=T(T-1) \ , \\
{\wtilde T}_{0}=T \ , \\
{\wtilde R}_0=R\ , \qquad (R=2I^0-(T-3/2)) \\
{\hat {\mib I}}^2=I(I+1) \ , \\
{\hat I}_0=I_0 \ , \\
{\hat M}_0=3I^1-M^0\ . \qquad (M^0=2(T-3/2)-I^0)
\end{cases}
\end{equation}
Further, we have 
\begin{equation}\label{4-7}
{\wtilde T}_-\ket{I^1I^0,II_0;T}=0 \ . 
\end{equation}
For the proof of the last relation in Eq.(\ref{4-5}), 
we used 
\b\label{4-8}
[\ {\hat M}_0 \ , \ {\hat D}_{I^1I_0^1}^* \ ]=3I^1{\hat D}_{I^1I_0^1}^* \ .
\end{equation}
Of course, the state $\ket{I^1I^0,II_0;T}$ is the eigenstate of 
${\hat \Gamma}_{su(3)}$ and $(I^1,I^0,I)$ obeys 
\b\label{4-9}
|I^1-I^0| \leq I \leq I^1+I^0 \ . 
\end{equation}
If the discussion is restricted only to the $su(3)$-algebra, 
it is enough to take into account the set $\{\ket{I^1I^0,II_0;T}\}$. 
If we include the $su(1,1)$-algebra, the state $\ket{I^1I^0,II_0;T}$ 
is generalized to the eigenstate of ${\wtilde T}_0$, the eigenvalue 
of which is different from $T$. 
The relation (\ref{4-7}) leads us to 
\beqn\label{4-10}
\ket{I^1I^0,II_0;TT^0}
&=&\sqrt{\frac{(2T-3)!}{(T_0-T)!(T_0+T-3)!}}({\wtilde T}_+)^{T_0-T}
\ket{I^1I^0,II_0;T} \ , \nonumber\\
& &\qquad\qquad
T_0=T\ , \ T+1\ , \ T+2 \ , \cdots \ . \quad (T_0\geq T)
\eeqn
Following an idea developed in the above, we have the orthogonal set for the 
$su(3)$-algebra and its associating $su(1,1)$-algebra.

\section{An orthogonal set apparently different from the form in \S 4}

As is clear from the treatment presented in \S 4, the derivation of the 
eigenstate (\ref{4-10}) is based on the fact that $({\hat D}_\pm^*)$ 
forms a spherical tensor with rank 1/2. 
In this section, we construct an orthogonal set in the framework of 
a set of two excited state generating operators which does not form 
the spherical tensor. 
It may be characteristic that, in contrast to the form obtained in \S 4, 
the form presented in this section is monomial. 

First, we introduce the following operators:
\b\label{5-1}
{\mib D}_-^*={\hat D}_-^* \ , \qquad
{\mib D}_+^*={\hat D}_+^*\cdot({\hat I}_0-1/2)
+(1/2){\hat I}_+\cdot{\hat D}_-^* \ . 
\end{equation}
They obey the relations 
\bsub\label{5-2}
\beqn
& &[ \ {\hat I}_- \ , \ {\mib D}_-^* \ ]=0 \ , \qquad 
[ \ {\hat I}_0 \ , \ {\mib D}_-^* \ ]=-(1/2){\mib D}_-^* \ , 
\label{5-2a}\\
& &[ \ {\hat I}_- \ , \ {\mib D}_+^* \ ]={\mib D}_+^*\cdot{\hat I}_- \ , 
\qquad
[ \ {\hat I}_0 \ , \ {\mib D}_+^* \ ]=(1/2){\mib D}_+^* \ , 
\label{5-2b}
\eeqn
\esub
\vspace{-0.8cm}
\beqn
& &[ \ {\hat M}_0 \ , \ {\mib D}_\pm^* \ ]=(3/2){\mib D}_{\pm}^* \ , 
\label{5-3}\\
& &[ \ {\mib D}_+^* \ , \ {\mib D}_-^* \ ]=0 \ , 
\label{5-4}\\
& &{\mib D}_-\ket{I^0,T}=0 \ , \qquad {\mib D}_+\ket{I^0,T}=0 \ . 
\label{5-5}
\qquad\qquad\qquad\quad
\eeqn
With the use of the operators ${\mib D}_{\pm}^*$, we introduce the 
following state:
\b\label{5-6}
\dket{I^1I^0,I;T}=({\mib D}_-^*)^{I^1-I^0+I}({\mib D}_+^*)^{I^1+I^0-I}
\ket{I^0,T} \ .
\end{equation}
Here, the normalization constant is omitted. 
The state (\ref{5-6}) obeys the following relations:
\bsub\label{5-7}
\beqn
& &{\wtilde T}_-\dket{I^1I^0,I;T}=0 \ , 
\label{5-7a}\\
& &{\wtilde T}_0\dket{I^1I^0,I;T}=T\dket{I^1I^0,I;T} \ , 
\label{5-7b}\\
& &{\wtilde R}_0\dket{I^1I^0,I;T}=R\dket{I^1I^0,I;T} \ , 
\qquad\qquad\quad
\label{5-7c}
\eeqn
\esub
\vspace{-0.9cm}
\bsub\label{5-8}
\beqn
& &{\hat I}_-\dket{I^1I^0,I;T}=0 \ , 
\label{5-8a}\\
& &{\hat I}_0\dket{I^1I^0,I;T}=-I\dket{I^1I^0,I;T} \ , 
\label{5-8b}\\
& &{\hat M}_0\dket{I^1I^0,I;T}=(3I^1-M^0)\dket{I^1I^0,I;T} \ . 
\label{5-8c}
\eeqn
\esub
Here, $R$ and $M^0$ are given in the relations (\ref{4-5}).

The relations (\ref{5-7}) and (\ref{5-8}) permit us to introduce the state 
$\dket{I^1I^0,II_0;TT_0}$ in the form 
\b\label{5-9}
\dket{I^1I^0,II_0;TT_0}=({\wtilde T}_+)^{T_0-T}({\hat I}_+)^{I+I_0}
\dket{I^1I^0,I;T} \ . 
\end{equation}
The state (\ref{5-9}) is the eigenstate of ${\wtilde {\mib T}}^2$, 
${\wtilde T}_0$, ${\wtilde R}_0$, ${\hat {\mib I}}^2$, ${\hat I}_0$ 
and ${\hat M}_0$ with the eigenvalues $T(T-1)$, $T_0$, $R$, $I(I+1)$, 
$I_0$ and $3I^1-M^0$, respectively. 
The exponents of ${\mib D}_{\pm}^*$ appearing in the state (\ref{5-6}) 
should be positive integers: 
\b\label{5-10}
I^1-I^0+I \geq 0 \ , \qquad I^1+I^0-I \geq 0 \ . 
\end{equation}
Further, we note the relation 
\beqn\label{5-11}
& &{\hat I}_-({\mib D}_+^*)^{I^1+I^0-I}\ket{I^0,T}=0 \ , \nonumber\\
& &{\hat I}_0({\mib D}_+^*)^{I^1+I^0-I}\ket{I^0,T}
=\left[(1/2)(I^1+I^0-I)-I^0\right]
({\mib D}_+^*)^{I^1+I^0-I}\ket{I^0,T} \ . 
\eeqn
Since $(1/2)(I^1+I^0-I)-I^0=(1/2)(I^1-I^0-I)$, the state 
$({\mib D}_+^*)^{I^1+I^0-I}\ket{I^0,T}$ can be regarded as the state 
$\ket{(1/2)(I+I^0-I^1),T}$. 
Therefore, $(1/2)(I+I^0-I^1)$ should be positive integer:
\b\label{5-12}
I+I^0-I^1 \geq 0 \ . 
\end{equation}
Combining the relations (\ref{5-10}) and (\ref{5-12}), we have 
\b\label{5-13}
|I^1-I^0|\leq I \leq I^1+I^0 \ . 
\end{equation}
Thus, we can learn that the state $\dket{I^1I^0,II_0;TT_0}$ is 
identical to the state $\ket{I^1I^0,II_0;TT_0}$ defined in the 
relation (\ref{4-10}). 
Of course, the normalization constant is omitted. 
The state (\ref{5-9}) is monomial, i.e., the state (\ref{5-9}) is 
obtained by successive operation of ${\wtilde T}_+$, ${\hat I}_+$, 
${\mib D}_-^*$ and ${\mib D}_+^*$ on the state $\ket{I^0,T}$.

\section{Discussion}

In \S\S 4 and 5, we presented the eigenvalue problem related to the 
$su(3)$-algebra and its associating $su(1,1)$-algebra. 
In this section, mainly, we discuss the case of the symmetric representation. 
The intrinsic state $\ket{m}$ in the present case can be expressed as 
\b\label{6-1}
\ket{m}=\ket{m_0,m_1}=\left(\sqrt{m_0!m_1!}\right)^{-1}
({\hat b}_-^*)^{m_1}({\hat b}^*)^{m_0}\ket{0} \ . 
\end{equation}
Here, $m_0$ and $m_1$, which are used in (A), are given as 
\b\label{6-2}
m_0=T-3/2-R=2[(T-3/2)-I^0] \ , \qquad 
m_1=T-3/2+R=2I^0 \ . 
\end{equation}
In (A), we showed that the cases $m_1=0$ and $m_0=0$ correspond to 
the symmetric representation, respectively. 
In this section, we discuss these cases in more detail. 
For this purpose, we note the relation 
\bsub\label{6-3}
\beqn
& &({\wtilde T}_0-3/2-{\wtilde R}_0)\ket{I^1I^0,II_0;T}
=m_0\ket{I^1I^0,II_0;T} \ , 
\label{6-3a}\\
& &({\wtilde T}_0-3/2+{\wtilde R}_0)\ket{I^1I^0,II_0;T}
=m_1\ket{I^1I^0,II_0;T} \ , 
\label{6-3b}
\eeqn
\esub
The above is derived from the relations (\ref{3-3}) and (\ref{6-2}). 
The relations (\ref{2-11}), (\ref{2-14}) and (\ref{6-3}) lead us to 
\bsub\label{6-4}
\beqn
& &({\hat b}^*{\hat b}+{\hat a}_+^*{\hat a}_+ +{\hat a}_-^*{\hat a}_-)
\ket{I^1I^0,II_0;T}
=m_0\ket{I^1I^0,II_0;T} \ , 
\label{6-4a}\\
& &({\hat a}^*{\hat a}+{\hat b}_+^*{\hat b}_+ +{\hat b}_-^*{\hat b}_-)
\ket{I^1I^0,II_0;T}
=m_1\ket{I^1I^0,II_0;T} \ , 
\label{6-4b}
\eeqn
\esub

Let us start in the case $m_1=0$. 
In this case, the relation (\ref{6-2}) tells us that $I^0=0$, and then, 
$I^1=I$: 
$\ket{I0,II_0;T}$. 
Since $m_1=0$, the relation (\ref{6-4b}) is equivalent to 
\b\label{6-5}
{\hat a}^*{\hat a}\ket{I0;II_0;T}=
{\hat b}_+^*{\hat b}_+\ket{I0;II_0;T}=
{\hat b}_-^*{\hat b}_-\ket{I0;II_0;T}=0 \ . 
\end{equation}
The relation (\ref{6-5}) shows that the state $\ket{I0,II_0;T}$ does 
not contain ${\hat a}^*$, ${\hat b}_+^*$ and ${\hat b}_-^*$. 
Therefore, if the treatment is restricted to the set 
$\{\ket{I0,II_0;T}\}$, which is a subspace of the space 
$\{\ket{II^0,II_0;TT_0}\}$, it is not necessary to take into account 
the bosons $({\hat a},{\hat a}^*)$, $({\hat b}_+,{\hat b}_+^*)$ and 
$({\hat b}_-,{\hat b}_-^*)$, explicitly. 
Then, omitting these boson operators, the $su(3)$-generators are expressed 
in the form 
\bsub\label{6-6}
\beqn
& &{\hat I}_+={\hat a}_+^*{\hat a}_- \ , \qquad
{\hat I}_-={\hat a}_-^*{\hat a}_+ \ , \qquad
{\hat I}_0=(1/2)({\hat a}_+^*{\hat a}_+-{\hat a}_-^*{\hat a}_-) \ , 
\label{6-6a}\\
& &{\hat M}_0=-{\hat b}^*{\hat b}+(1/2)({\hat a}_+^*{\hat a}_+
+{\hat a}_-^*{\hat a}_-) \ , 
\label{6-6b}\\
& &{\hat D}_-^*={\hat a}_-^*{\hat b}\ , \qquad
{\hat D}_+^*={\hat a}_+^*{\hat b} \ , \qquad
{\hat D}_-={\hat b}^*{\hat a}_- \ , \qquad
{\hat D}_+={\hat b}^*{\hat a}_+ \ . 
\label{6-6c}
\eeqn
\esub
The operators $({\wtilde T}_{\pm,0})$ and ${\wtilde R}_0$ are reduced to 
\bsub\label{6-7}
\beqn
& &{\wtilde T}_{\pm}=0 \ , 
\label{6-7a}\\
& &{\wtilde T}_0-3/2=-{\wtilde R}_0
=(1/2)({\hat b}^*{\hat b}+{\hat a}_+^*{\hat a}_++{\hat a}_-^*{\hat a}_-) \ . 
\label{6-7b}
\eeqn
\esub
We can learn that the form (\ref{6-6}) is identical with the 
symmetric representation.

Next, we consider the case $m_0=0$. 
As was discussed briefly in (A), this case essentially belongs to 
the symmetric representation. 
In order to show this fact explicitly, we relabel the boson 
operators as follows: 
\beqn\label{6-8}
& &{\hat a} \longrightarrow {\hat a}_+ \ , \qquad
{\hat a}_- \longrightarrow {\hat b}_+ \ , \qquad
{\hat a}_+ \longrightarrow {\hat a} \ , \nonumber\\
& &{\hat b} \longrightarrow {\hat b}_- \ , \qquad
{\hat b}_- \longrightarrow {\hat b} \ , \qquad
{\hat b}_+ \longrightarrow -{\hat a}_- \ . 
\eeqn
This procedure was sketched in (A) in different notations. 
Then, the $su(3)$-generators shown in the form (\ref{2-10}) are relabeled as 
\bsub\label{6-8add}
\beqn
& &{\hat I}_+ \longrightarrow {\hat D}_-^* \ , \qquad
{\hat I}_- \longrightarrow {\hat D}_- \ , \qquad
{\hat I}_0 \longrightarrow (1/2)({\hat M}_0-{\hat I}_0) \ , 
\label{6-8a}\\
& &{\hat M}_0 \longrightarrow (1/2)({\hat M}_0+3{\hat I}_0) \ , 
\label{6-8b}\\
& &{\hat D}_-^* \longrightarrow -{\hat I}_+ \ , \qquad
{\hat D}_+^* \longrightarrow {\hat D}_+^* \ , \qquad
{\hat D}_- \longrightarrow -{\hat I}_- \ , \qquad
{\hat D}_+ \longrightarrow {\hat D}_+ \ . 
\label{6-8c}
\eeqn
\esub
The $su(1,1)$-generators and ${\wtilde R}_0$ are relabeled as 
\beqn
& &{\wtilde T}_{\pm} \longrightarrow -{\wtilde T}_{\pm} \ , \qquad
{\wtilde T}_0 \longrightarrow {\wtilde T}_0 \ , 
\label{6-9}\\
& &{\wtilde R}_0 \longrightarrow -{\wtilde R}_0 \ . 
\label{6-10}
\eeqn
Therefore, we can conclude that under the above relabeling, the case 
$m_0=0$ is reduced to the case $m_1=0$ which was already discussed. 

In the next paper, on the basis of the relation (\ref{6-4}), 
we will present the Holstein-Primakoff boson realization.

\section*{Acknowledgements} 

The work presented in the series of papers, in which the present 
is the first, has been initiated in summer of 2004 and finished in 
summer of 2005. 
In these periods, the authors, Y. T. and M. Y. have stayed at Coimbra 
under the invitaion by Professor J. da Providencia, the co-author 
of this paper. 
They should acknowledge not only for his kind and repeated invitations 
but also for his many and valuable suggestions. 



\appendix
\section{Third form of the orthogonal set}

In this Appendix, we show the third form of the orthogonal set. 
The first and the second forms are shown in \S 4 and \S 5, respectively. 
For the preparation, we consider the following operator: 
\beqn\label{a-1}
& &{\hat P}=1-{\hat I}_+({\hat I}_-{\hat I}_+ + \epsilon)^{-1}{\hat I}_- \  ,
\qquad 
{\hat I}_-{\hat I}_+={\hat {\mib I}}^2-{\hat I}_0({\hat I}_0+1) \ , 
\nonumber\\
& &\epsilon\ :\ \hbox{\rm infinitesimal\ parameter}\ . 
\eeqn
Here, $({\hat I}_{\pm,0})$ obeys the $su(2)$-algebra, for example, 
shown in the relation (\ref{2-10a}). 
In order to avoid null denominator, $\epsilon$ is introduced. 
Let $\ket{I,I_0}$ be the eigenstate of ${\hat {\mib I}}^2$ and ${\hat I}_0$ 
with the eigenvalues $I(I+1)$ and $I_0$, respectively. 
Then, we have the following relation: 
\beqn\label{a-2}
{\hat P}\ket{I,I_0}&=&\epsilon\left(
(I+I_0)(I-I_0+1)+\epsilon\right)^{-1}\ket{I,I_0} \nonumber\\
&\stackrel{\epsilon\rightarrow 0}{\longrightarrow}&
\begin{cases}\ket{I,-I} \ , & (I_0=-I) \\
0 \ . & (I_0=-I+1\ , \ -I+2\ , \ \cdots \ , \ I)
\end{cases}
\eeqn
Therefore, for an appropriately normalized eigenstate of 
${\hat I}_0$, $\ket{I}$, we have 
\begin{equation}\label{a-3}
{\hat P}\ket{I}=\ket{I,-I} \ . 
\end{equation}
We can see that the operator ${\hat P}$ at the limit $\epsilon \rightarrow 0$ 
plays a role of the projection operator for pick up $\ket{I,-I}$ 
from $\ket{I}$.

Under the above preparation, we consider the $su(3)$-algebra. 
First, we note the relations 
\b\label{a-4}
[\ {\wtilde T}_{\pm,0} \ , \ {\hat P} \ ]=0 \ , \qquad
[\ {\wtilde R}_0 \ , \ {\hat P} \ ]=0 \ , 
\end{equation}
\vspace{-0.8cm}
\bsub
\beqn
& &{\hat I}_-{\hat P}=\epsilon\cdot({\hat I}_-{\hat I}_+
+\epsilon)^{-1}{\hat I}_- \ , 
\label{a-5a}\\
& &{\hat P}{\hat I}_+=\epsilon\cdot {\hat I}_+({\hat I}_-{\hat I}_+
+\epsilon)^{-1} \ , 
\label{a-5b}\\
& &[\ {\hat I}_0 \ , \ {\hat P}\ ]=0 \ , 
\label{a-5c}
\eeqn
\esub
\vspace{-0.9cm}
\b\label{a-6}
[ \ {\hat M}_0 \ , \ {\hat P}\ ]=0 \ . \qquad\qquad\quad
\end{equation}
Let us investigate the state $\rket{I^1I^0,I;T}$ which is 
obtained by replacing $\mib{D}_-^*$ and ${\mib D}_+^*$ with 
${\hat D}_-^*$ and ${\hat D}_+^*$, respectively, in the state (\ref{5-6}): 
\b\label{a-7}
\rket{I^1I^0,I;T}=({\hat D}_-^*)^{I^1-I^0+I}({\hat D}_+^*)^{I^1+I^0-I}
\ket{I^0,T} \ . 
\end{equation}
The state (\ref{a-7}) satisfies the same relations as those shown in 
the relations (\ref{5-7a})$\sim$(\ref{5-8c}) except (\ref{5-8a}):
\b\label{a-8}
{\hat I}_-\rket{I^1I^0,I;T} \neq 0 \ . 
\end{equation}
Then, we define the state 
\b\label{a-9}
\rdket{I^1I^0,I;T}={\hat P}\rket{I^1I^0,I;T} \ . 
\end{equation}
The relations (\ref{a-3}) and (\ref{a-5a}) give us 
\b\label{a-10}
{\hat I}_-\rdket{I^1I^0,I;T}
\stackrel{\epsilon\rightarrow 0}{\longrightarrow} 0 .
\end{equation}
With the use of the relations (\ref{a-4})$\sim$(\ref{a-6}), we can 
prove the relations (\ref{5-7a})$\sim$(\ref{5-8c}). 
Of course, $\{\rdket{I^1I^0,I;T}\}$ is orthogonal. 
Then, in the same form as that shown in the relation (\ref{5-9}), 
we have 
\b\label{a-11}
\rdket{I^1I^0,II_0;TT_0}=({\wtilde T}_+)^{T_0-T}
({\hat I}_+)^{I+I_0}\rdket{I^1I^0,I;T} \ . 
\end{equation}
We can derive the state (\ref{a-11}) by successive operation of 
${\wtilde T}_+$, ${\hat I}_+$, ${\hat P}$, ${\hat D}_-^*$ and 
${\hat D}_+^*$. 
The above is the third form for the orthogonal set of the 
$su(3)$-algebra.

\end{document}